# AN IMPROVED VARIABLE STEP-SIZE AFFINE PROJECTION SIGN ALGORITHM FOR ECHO CANCELLATION[*]

*Jianming Liu and Steven L Grant*[1]

Department of Electrical and Computer Engineering, Missouri University of Science and Technology, Rolla, Missouri 65409.

## ABSTRACT

This paper proposes an improved variable step-size (VSS) algorithm for the recently introduced affine projection sign algorithm (APSA) based on the recovery of the near-end signal energy in the error signal. Simulation results demonstrate that, compared to the previous VSS for APSA, the proposed approach provides both more robustness to impulse interference and better tracking ability of echo path change.

## 1. INTRODUCTION

The problem of acoustic echo cancellation (AEC) is usually approached by modeling the echo path impulse with an adaptive filter and subtracting the estimated echo from the microphone output signal [1]. However, due to the near-end signal and echo path change, it is very important that the adaptive filter algorithm should be robust to impulsive interference and sensitive to echo path change.

Recently, for acoustic echo cancellation, the family of sign algorithms has become popular in the literature because they are robust to impulsive interference [2]-[3]. It is well known that the performance of most adaptive filters is governed by the choice of the step-size parameter [4]-[6]. A few variable step-size algorithms for sign algorithms have been reported in [7]-[10]. These variable step-size algorithms, however, are not robust to impulsive interference or sensitive to echo path change. The variable step-size algorithm in [10] depends on a priori information of the interference. Meanwhile, VSS in [7] and the delta sequence in [8]-[9] could not react to the change of echo path. In this paper, we propose a variable step-size algorithm the affine projection sign algorithm. Meanwhile, the proposed VSS APSA algorithm is robust to impulsive interference and tracks echo path changes quickly.

This paper is organized as follows. Section 2 reviews the recently proposed VSS for APSA, and in Section 3 we present the proposed VSS-APSA. The simulation results and comparison to the previous algorithms are presented in Section 4. Finally conclusions are drawn in Section 5.

## 2. REVIEW OF VSS APSA

The far-end signal $x(n)$ is filtered through the room impulse response $h(n)$ to get the echo signal $y(n)$.

$$y(n) = x(n) * h(n) = x_n^T h_n \qquad (1)$$

where
$x_n = [x(n)\, x(n-1)\cdots x(n-L+1)]^T$, $h_n = [h_0\, h_1 \cdots h_{L-1}]^T$, and $L$ is the length of echo path. This echo signal is added to the near-end signal $v(n)$ (including both speech and back-ground noise, etc.) to get the microphone signal $d(n)$,

$$\begin{aligned} d(n) &= x(n) * h(n) + v(n) \\ &= x_n^T h_n + v(n). \end{aligned} \qquad (2)$$

Grouping the $P$ most recent input vectors $x(n)$ together gives the input signal matrix:

$$X(n) = [x(n)\, x(n-1)\cdots x(n-P+1)]$$

We define the *a priori* and *a posteriori* error vectors as

$$e(n) = d(n) - X^T(n)\hat{h}(n-1), \text{ and} \qquad (3)$$

$$\varepsilon(n) = d(n) - X^T(n)\hat{h}(n). \qquad (4)$$

This error, $e(n)$ is used to adapt the AEC filter $\hat{h}(n)$. The APSA algorithm updates the filter coefficients as follows:

$$\hat{h}(n) = \hat{h}(n-1) - \mu(n)\frac{X(n)\operatorname{sgn}(e(n))}{\sqrt{\operatorname{sgn}(e^T(n))X^T(n)X(n)\operatorname{sgn}(e(n))}}. \qquad (5)$$

in which $\mu(n)$ is the variable step-size. In [7], a variable step-size was proposed for APSA as follows:

$$\mu(n) = \alpha\mu(n-1) + (1-\alpha)\min\left(\frac{\|e(n)\|_1}{\sqrt{\text{sgn}(e^T(n))X^T(n)X(n)\text{sgn}(e(n))}}, \mu(n-1)\right). \quad (6)$$

It should be noted that, in the variable step-size (6) uses the absolute value of error signal which includes the noise. However, this will cause larger step-size and higher misalignment at steady-state. Meanwhile, (6) is very similar to the delta sequence in [8]-[9], which has the decreasing property. Although the algorithm becomes more robust against perturbations, it also loses its tracking capacity. Unfortunately, this is not practical when echo path changes are possible and additional *ad hoc* control has to be included, which increases the complexity of control logic.

Meanwhile, in [10], another variable step-size was proposed for normalized sign algorithm as follows

$$\mu(n) = \frac{E[|e(n)|] - E[|v(n)|]}{\sqrt{x^T(n)x(n)}} \quad (7)$$

Although (7) was not proposed for affine projection sign algorithm, we will show that this could be extended to affine projection sign algorithm in Section 3. However, the main disadvantage is that it suffers from requiring the estimation of $E[|v(n)|]$, since the near-end signal is buried by residual error signal. Therefore, this algorithm depends on the priori information about the near-end signal.

Compared to previous VSS in [7]-[10], our proposed algorithm overcomes the above problems through the estimation of the near-end signal energy and provides both robustness to impulsive interference and tracking ability of echo path change.

## 3. PROPOSED VSS-APSA

We will derive our proposed variable step-size algorithm as follows. Similar to [6], we rewrite the update of APSA in (5) as the following form:

$$\hat{h}(n) = \hat{h}(n-1) - \frac{X(n)\mu(n)\text{sgn}(e(n))}{\sqrt{\text{sgn}(e^T(n))X^T(n)X(n)\text{sgn}(e(n))}}. \quad (8)$$

where

$$\mu(n) = diag\{\mu_0(n), \mu_1(n), \ldots, \mu_{P-1}(n)\} \quad (9)$$

is a $P \times P$ diagonal matrix. It is obvious that the normal APSA is obtained when $\mu_0(n) = \mu_1(n) = \cdots = \mu_{P-1}(n) = \mu$. Substituting (8) into (3) and (4), we have

$$\varepsilon(n) = e(n) - \frac{X^T(n)X(n)\mu(n)\text{sgn}(e(n))}{\sqrt{\text{sgn}(e^T(n))X^T(n)X(n)\text{sgn}(e(n))}} \quad (10)$$

According to [5] and [6], acoustic echo cancellation can be viewed as the recovery of the "useful" signal (i.e., the near-end signal) from the error signal of adaptive filter. Therefore, a more reasonable target should be $\varepsilon(n) = v(n)$, where $v(n) = [v(n), v(n-1), \cdots, v(n-P+1)]^T$ is the near-end signal vector.

In order to simplify the analysis, we use the following diagonal assumption:
$$X^T(n)X(n) = diag\{x^T(n)x(n), x^T(n-1)x(n-1), \ldots, x^T(n-P+1)x(n-P+1)\}.$$

and define the symbol $\delta_n$ as

$$\delta_n = \sqrt{\text{sgn}(e^T(n))X^T(n)X(n)\text{sgn}(e(n))}.$$

Therefore, we have

$$\varepsilon_{l+1}(n) = e_{l+1}(n) - \frac{1}{\delta_n}\mu_l(n)x^T(n-l)x(n-l)\text{sgn}(e_{l+1}(n)). \quad (11)$$

where the variables $\varepsilon_{l+1}(n)$ and $e_{l+1}(n)$ denote the $(l+1)^{th}$ elements of the vectors $\varepsilon(n)$ and $e(n)$, $l = 0,1,\ldots,P-1$. We can generalize the variable step-size in [10] to APSA based on the following condition.

$$E[|\varepsilon_{l+1}(n)|] = E[|v_{l+1}(n)|]. \quad (12)$$

The variable step-size is then seen to be

$$\mu_l(n) = \frac{\delta_n(E[|e_{l+1}(n)|] - E[|v_{l+1}(n)|])}{E[x^T(n-l)x(n-l)]} \quad (13)$$

However, as mentioned before, it is difficult to estimate $E[|v_{l+1}(n)|]$ in real time, and in order to overcome this difficulty, we propose to use the following criterion as in [5] and [6].

$$E[\varepsilon_{l+1}^2(n)] = E[v_{l+1}^2(n)] \quad (14)$$

Squaring (11) and taking expectations result in

$$\mu_l^2 \frac{E\left[(x_{l+1}^T x_{l+1})^2\right]}{\delta_n^2} - \mu_l \frac{2E[x_{l+1}^T x_{l+1}|e_{l+1}|]}{\delta_n} + E[e_{l+1}^2] - E[v_{l+1}^2] = 0. \quad (15)$$

in which we denote $\boldsymbol{x}_{l+1} = \boldsymbol{x}_{l+1}(n) = \boldsymbol{x}(n-l)$ for brevity. Solving this quadratic equation we get:

$$\mu_l(n) = \frac{\delta_n}{E\left[\left(\boldsymbol{x}_{l+1}^T \boldsymbol{x}_{l+1}\right)^2\right]} \left[ E\left[\boldsymbol{x}_{l+1}^T \boldsymbol{x}_{l+1} |e_{l+1}|\right] - \sqrt{\left(E\left[\boldsymbol{x}_{l+1}^T \boldsymbol{x}_{l+1} |e_{l+1}|\right]\right)^2 - E\left[\left(\boldsymbol{x}_{l+1}^T \boldsymbol{x}_{l+1}\right)^2\right]\left(E\left[e_{l+1}^2\right] - E\left[v_{l+1}^2\right]\right)} \right]. \quad (16)$$

where we have kept the smaller solution as the step-size to ensure stability.

The energy of the near-end signal $E\left[v_{l+1}^2\right]$ in (16) is still not directly accessible, therefore similar to the near-end signal energy estimator in [11], we propose to estimate $E\left[e_{l+1}^2\right] - E\left[v_{l+1}^2\right]$ for APSA as follows.

$$E\left[e_{l+1}^2(n)\right] - E\left[v_{l+1}^2(n)\right] = \frac{\boldsymbol{r}_{xe,l+1}^T \boldsymbol{r}_{x\mathrm{sgn}e,l+1}}{E\left[\sqrt{\boldsymbol{x}_{l+1}^T(n) \boldsymbol{x}_{l+1}(n)}\right]} \quad (17)$$

in which

$$\boldsymbol{r}_{x\mathrm{sgn}e,l+1} = E\left[\boldsymbol{x}_{l+1}^T(n) \mathrm{sgn}(e_{l+1}(n))\right] \quad (18)$$

$$\boldsymbol{r}_{xe,l+1} = E\left[\boldsymbol{x}_{l+1}^T(n) e_{l+1}(n)\right]. \quad (19)$$

Therefore, the proposed variable step-size for APSA is:

$$\mu_l(n) = \frac{\delta_n}{E\left[\left(\boldsymbol{x}_{l+1}^T \boldsymbol{x}_{l+1}\right)^2\right]} \left[ E\left[\boldsymbol{x}_{l+1}^T \boldsymbol{x}_{l+1} |e_{l+1}|\right] - \sqrt{\left(E\left[\boldsymbol{x}_{l+1}^T \boldsymbol{x}_{l+1} |e_{l+1}|\right]\right)^2 - \frac{E\left[\left(\boldsymbol{x}_{l+1}^T \boldsymbol{x}_{l+1}\right)^2\right] \boldsymbol{r}_{xe,l+1}^T \boldsymbol{r}_{x\mathrm{sgn}e,l+1}}{E\left[\sqrt{\boldsymbol{x}_{l+1}^T(n) \boldsymbol{x}_{l+1}(n)}\right]}} \right] \quad (20)$$

Meanwhile, for practical implementation, we compute the expectations in the following recursive manner.

$$\hat{\sigma}_{x|e|,l+1}(n+1) = E\left[\boldsymbol{x}_{l+1}^T \boldsymbol{x}_{l+1} |e_{l+1}|\right] = \alpha \hat{\sigma}_{x|e|,l+1}(n) + (1-\alpha)\boldsymbol{x}^T(n-l)\boldsymbol{x}(n-l)|e_{l+1}(n)|, \quad (21)$$

$$\hat{\sigma}_{x^2,l+1}(n+1) = E\left[\left(\boldsymbol{x}_{l+1}^T \boldsymbol{x}_{l+1}\right)^2\right] = \alpha \hat{\sigma}_{x^2,l+1}(n) + (1-\alpha)\left(\boldsymbol{x}^T(n-l)\boldsymbol{x}(n-l)\right)^2, \quad (22)$$

$$\hat{\sigma}_{\sqrt{x},l+1}(n+1) = E\left[\sqrt{\boldsymbol{x}^T(n-l)\boldsymbol{x}(n-l)}\right] = \alpha \hat{\sigma}_{\sqrt{x},l+1}(n) + (1-\alpha)\sqrt{\boldsymbol{x}^T(n-l)\boldsymbol{x}(n-l)}, \quad (23)$$

$$\hat{\boldsymbol{r}}_{xe,l+1}(n+1) = \alpha \hat{\boldsymbol{r}}_{xe,l+1}(n) + (1-\alpha)\boldsymbol{x}(n-l)e_{l+1}(n), \quad (24)$$

$$\hat{\boldsymbol{r}}_{x\mathrm{sgn}e,l+1}(n+1) = \alpha \hat{\boldsymbol{r}}_{x\mathrm{sgn}e,l+1}(n) + (1-\alpha)\boldsymbol{x}(n-l)\mathrm{sgn}(e_{l+1}(n)). \quad (25)$$

where $\alpha$ is a smoothing factor.

Meanwhile, in order to further smooth the proposed variable step-size and avoid calculating the square root of negative number, we propose to use the following smoothed step-size in practice:

$$\mu_l(n) = \alpha \mu_l(n-1) + (1-\alpha)\delta_n \left[ \frac{\hat{\sigma}_{x|e|,l+1}(n)}{\hat{\sigma}_{x^2,l+1}(n)} - \sqrt{\max\left\{ \left(\frac{\hat{\sigma}_{x|e|,l+1}(n)}{\hat{\sigma}_{x^2,l+1}(n)}\right)^2 - \frac{\hat{\boldsymbol{r}}_{xe,l+1}^T \hat{\boldsymbol{r}}_{x\mathrm{sgn}e,l+1}}{\hat{\sigma}_{x^2,l+1}(n)\hat{\sigma}_{\sqrt{x},l+1}(n)}, 0 \right\}} \right] \quad (26)$$

## 4. SIMULATION RESULTS

We do computer simulations in the scenario of acoustic echo cancellation. We use a random echo path with length, $L=128$, and the adaptive filter is with the same length. The projection order for the affine projection sign algorithm is $P=5$.

The colored input signals are generated by filtering white Gaussian noise (WGN) through a first order system with a pole at 0.8. Independent white Gaussian noise is added to the system background with a signal-to-noise ratio, SNR = 30 dB. The impulsive noise is generated as a Bernoulli-Gaussian (BG) distribution with signal-to-interference ratio (SIR). The Bernoulli-Gaussian distribution was generated as a product of a Bernoulli process and a Gaussian process, i.e., $z(k) = w(k)n(k)$, where $n(k)$ was WGN with zero mean and variance $\sigma_n^2$, and $w(k)$ was a Bernoulli process with the probability mass function given as $P(w) = 1 - P_r$ for $w = 0$, and $P(w) = P_r$ for $w = 1$. The average power of the BG process was $P_r \sigma_n^2$ and in our simulation, the probability for Bernoulli process is 0.1 [3]. The convergence state of adaptive filter is evaluated with the normalized misalignment which is defined as $10\log_{10}(\|\boldsymbol{h} - \hat{\boldsymbol{h}}\|_2 / \|\boldsymbol{h}\|_2)$

For the VSS in [10], considering we are using colored input signals, the far-end and echo signal are always there, thus we could not estimate the priori information $E[|v(n)|]$ of near-end signal during the silence period of echo signal. In our simulation, we estimate $E[|v(n)|]$ directly from the already known background WGN and BG signals which are added in the microphone signal. However, since this near-end signal is not available in practice, we only take this as the theoretically optimal VSS we could obtain. Meanwhile, we also compare proposed VSS with two fixed step-size APSAs (0.01 and 0.001), one fixed step-size APA (0.1) and Shin's VSS APSA in [7].

At first, when there is no impulsive interference but only background noise with SNR =30 dB, we compare the performance of above algorithms as in Fig. 1(a). The parameters of VSS APSAs are chosen to allow all the VSS APSAs have both similar convergence rate and steady-state misalignment (about -25 dB) as APA with fixed step-size 0.1. The APSA with fixed step-size 0.01 is chosen because it has the similar convergence rate with VSS APSAs, and APSA with fixed step-size 0.001 will have about -25 dB steady-state misalignment too. We further demonstrate the variation of step-size in Fig. 1(b) at the same time. We could observe that the VSS APSAs provide a good trade-off between the convergence rate and steady-state misalignment compared with the two fixed step-size APSAs.

Secondly, when there are both strong BG impulsive interference with SIR = 0 dB and background noise with SNR = 30 dB, we compare the normalized misalignment and VSS in Fig. 2. We could clearly observe the advantage of APSAs over APA since APA will diverge due to the strong impulse noise. At the same time, Shin's VSS in [7] will have a higher steady-state misalignment since it involves the noise in the step-size at steady-state. However, the proposed VSS APSA has a lower misalignment at steady-state due to the recovery of the impulsive interference and almost approaches the theoretically optimal performance of Shao's algorithm in [10].

Finally, we will compare the tracking ability of echo path change for the above different algorithms as in Fig. 3. We simulate the echo path change at sample 10000 by switching to another random echo path, and results demonstrate that the proposed VSS APSA provides a good approximation of theoretical performance, which means good trade-off between fast tracking ability and lower steady-state misalignment.

To sum up, our improved variable step-size for APSA could almost approach the theoretically optimal performance of Shao's VSS algorithm and outperforms both the Shin's VSS and fixed step-size APSAs in terms of the convergence rate and steady-state misalignment.

## 5. CONCLUSION

We have proposed an improved variable step-size for the recently proposed affine projection sign algorithm based on the near-end signal energy recovery from the error signal. Simulation results demonstrate that our proposed algorithm is robust to impulsive interference and provides better tradeoff between lower steady-state misalignment and fast tracking ability for echo path change compared to the previous ones.

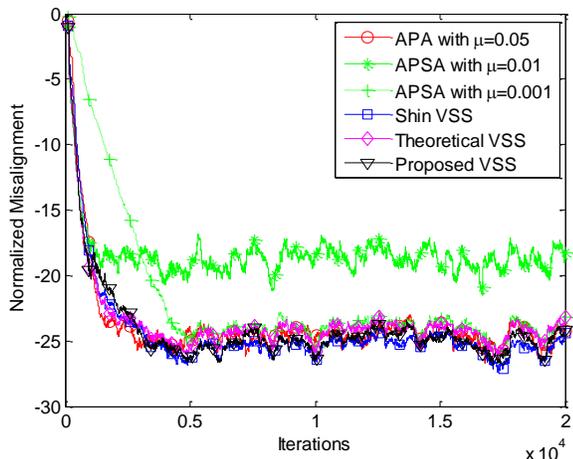

**Fig.1 (a)** Comparison of normalized misalignment with SNR=30dB.

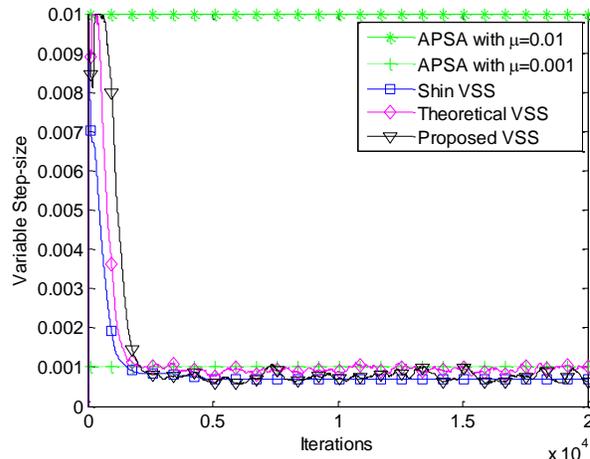

**Fig.1 (b)** Comparison of variable step-size with SNR=30dB.

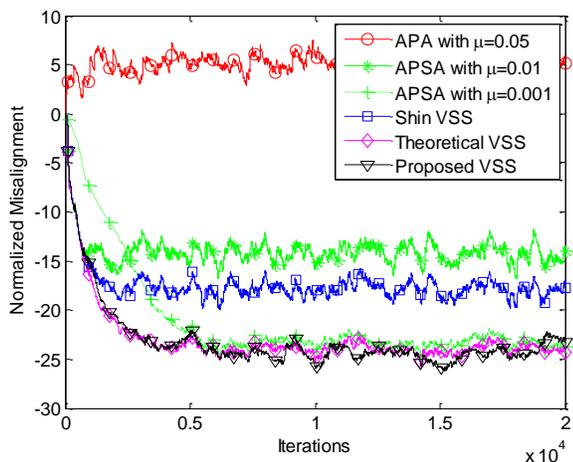

**Fig.2 (a)** Comparison of normalized misalignment with impulsive interference SIR=0dB and SNR=30dB.

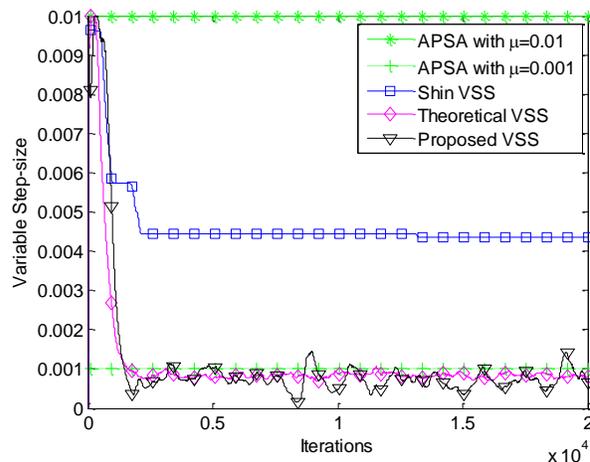

**Fig.2 (b)** Comparison of variable step-size with impulsive interference SIR=0dB and SNR=30dB.

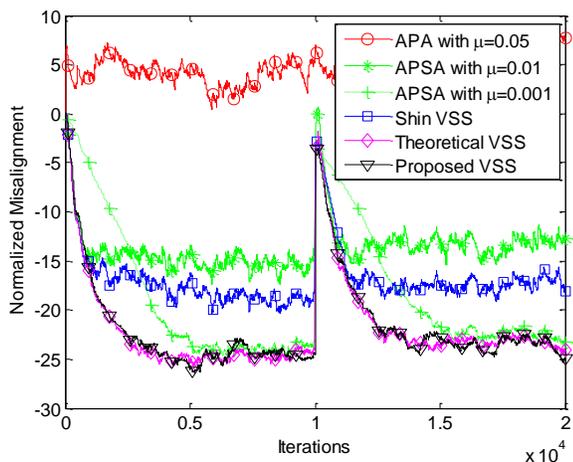

**Fig.3 (a)** Comparison of normalized misalignment with echo path change at 10000, SIR=0dB and SNR=30dB.

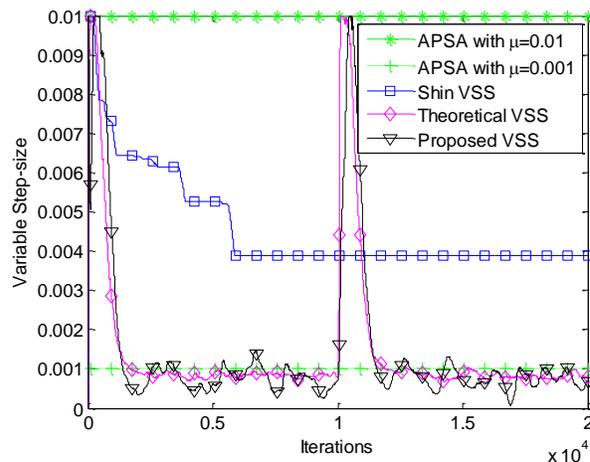

**Fig.3 (b)** Comparison of variable step-size with echo path change at 10000, SIR=0dB and SNR=30dB.